# Electro-Optic Co-Simulation in High-Speed Silicon Photonics Transceiver Design Using Standard Electronic Circuit Simulator

**KEISUKE KAWAHARA**[1] **(Graduate Student Member, IEEE),\* AND TOSHIHIKO BABA**[1] **(Fellow, IEEE)**

[1]Department of Mathematics, Physics, Electrical Engineering and Computer Science, Yokohama National University, Yokohama, 240-8501, Japan

CORRESPONDING AUTHOR: K. Kawahara (e-mail: keisuke@ieee.org).

This work was supported by JSPS KAKENHI Grant Number JP23KJ0988.

**ABSTRACT** The increasing demand for high-speed optical interconnects necessitates integrated photonic and electronic solutions. Electro-optic co-simulation is key to meeting these requirements, which works by importing interoperable photonic models into industry-standard electronic circuit simulators from Synopsys, Cadence, Keysight, and others. However, current interoperable photonic models cannot accurately predict performance and do not address terabit-class transceiver designs due to inadequate modeling of complex physical effects such as optical losses, back-reflection, nonlinearity, high-frequency response, noise, and manufacturing variations. Here, we present accurate and interoperable photonic models that agree well with experiments at symbol rates exceeding 50 Gbaud. The developed models include basic optical components with losses and reflections, two types of Mach-Zehnder modulators with validated high-frequency response, and testing equipment with associated noise. We built an optical link testbench on an industry-standard electronic circuit simulator and verified the model accuracy by comparing simulation and experiment up to 64 Gbaud. The results suggest that co-simulation will be a solid basis for advancing design of transceivers and other related applications in silicon photonics.

## I. INTRODUCTION

The exponential growth of demand in data centers and high-performance computers necessitates high-speed optical interconnects. Terabit Ethernet requires symbol rates exceeding 50 Gbaud and compact form factors to meet this demand [1]. Integrating photonics and electronics addresses these requirements by minimizing connection losses and reducing the footprint [2–8]. Due to the high cost of photomasks and long lead times in silicon (Si) photonics manufacturing, accurate electro-optic (EO) co-simulation is essential to predicting performance of designed devices. An efficient approach is to perform co-simulation in a unified environment, eliminating complex data exchange between photonics and electronics. This can be accomplished by implementing photonic device models in a standardized language, such as Verilog-A [9], and importing them into industry standard electronic circuit simulators such as Synopsys, Cadence, or Keysight. These models must capture a variety of physical effects such as optical losses, back reflection, nonlinearity, high-frequency response, noise and manufacturing variations, which need to be experimentally validated. In earlier works, equivalent baseband models have allowed optical propagation to be simulated in electronic circuit simulators [10–16]. Similar models have made progress, including the introduction of complex vector fitting to take account of the wavelength spectrum. [17–22]. However, these studies are not based on experimental results and hardly reflect the real device characteristics, making them far from practical design. Several studies have developed measurement-based models for microring modulators [23, 24], Mach-Zehnder modulators [25, 26], and photodetectors [27]. To our knowledge, one of the most advanced photonic models is included in GlobalFoundries' Fotonix technology, which contains basic photonic components and a microring modulator with thermal response [28]. Nevertheless, the modeling and experimental validation of the physical properties required for designing transceivers remain insufficient. In particular, no reported models have fully validated the high-frequency response while considering the noise of active components. Furthermore, the details of the reported models are often not disclosed, which is hindering the progress of this type of research.

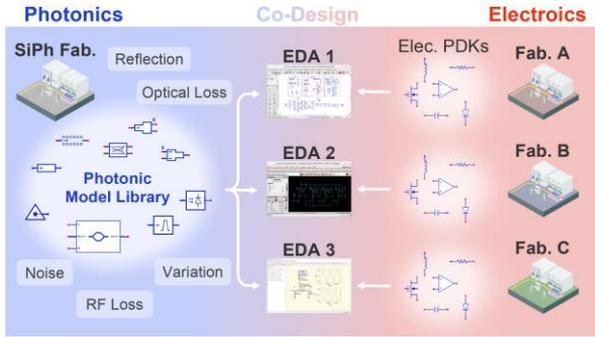

**FIGURE 1.** Concept of EO co-design using interoperable photonic device models and electronic circuit simulators. The model's interoperability allows any combination of photonics and electronics foundries.

In this study, we present validated and interoperable photonic device models that accurately predict transmission performance and support symbol rates in excess of 50 Gbaud, disclosing their details as well as source codes. Here, two types of Si Mach-Zehnder optical modulators were modeled, which include frequency-dependent lossy traveling-wave electrodes and slow-light enhancement of photonic crystal waveguides (PCW). In addition, models of test equipment with validated noise (erbium-doped fiber amplifier (EDFA), tunable filter, and photodetector module) were developed, and a test bench of full optical link was built. Our model library is interoperable between various simulator tools, thanks to the standardized syntax of Verilog-A, which allows for co-design with any combination of photonics and electronics foundries, as shown in Fig. 1.

In this paper, Section II presents the full process of Si photonics device modeling, and Section III describes the modeling of the test equipment for accurate noise level estimation. Section IV demonstrates the EO co-simulation with a developed testbench, validated up to 64 Gbaud. All models and sample testbenches implemented in Keysight's Advanced Desing System (ADS) are available on GitHub [29].

## II. SI PHOTONICS COMPONENTS

In this section, Si photonics component models are developed, targeting a CMOS-compatible 300-mm silicon-on-insulator wafer process by the National Institute of Advanced Industrial Science and Technology. The basic principle of the simulation is based on the equivalent baseband model [12]. Section III-A briefly reviews the simulation principles through a description of the waveguide model. Section III-B introduces the coupler models for optical splitting, combining, and loopback. Section III-C details two types of Si Mach-Zehnder modulator models with validated high-frequency responses and slow-light enhancement.

### A. Waveguide

The difficulty in introducing photonic devices into electronic circuit simulations is that the frequency of the optical carrier, $\omega_O$, is much higher than that of the electrical signal, requiring a very large sampling rate. For example, for a wavelength of $\lambda = 1550$ nm, $f = \omega_O/2\pi = 193$ THz. To address this difficulty, equivalent baseband simulations were introduced. Here, $\omega_O$ is shifted by a reference frequency, $\omega_{ref} = 2\pi c/\lambda_{ref}$. The electric field of the optical signal is expressed as

$$\tilde{E}_B(t) = \exp(-j\omega_{ref}t)\,\tilde{E}(t), \quad (1)$$

where $\tilde{E}(t)$ is the analytic signal [30] of the electric field $E(t)$. Typically, $\omega_{ref}$ is approximated by $\omega_O$. For example, the electric field of continuous wave laser light is represented as

$$\tilde{E}_{LD}(t) = A_{LD}\exp[j(\omega_o - \omega_{ref})t], \quad (2)$$

where $A_{LD}$ is the amplitude of the electrical field. To support complex numbers and bidirectional propagation in circuit simulators, bus wires $E[0:3]$ are used, where $E[0]$ and $E[1]$ represent the real and imaginary parts of forward propagation and $E[2]$ and $E[3]$ represent those of backward propagation, respectively. The propagation of light in a waveguide is modeled as

$$\tilde{E}_{out} = \exp\left[-\left(\frac{\alpha_0}{2} + \frac{j\omega_{ref}n_{eq}}{c}\right)l_{wg}\right]\tilde{E}_{in}\left(t - \frac{n_g}{c}l_{wg}\right), (3)$$

where $n_{eq}$, $\alpha_0$, $n_g$, and $l_{wg}$ represent modal equivalent refractive index, loss constant, group index, and waveguide length, respectively. A Si wire waveguide typically has $n_{eq} = 2.31$, $n_g = 4.34$, $a_0 = 2$dB/cm for the transverse-electric (TE) polarization at $\lambda = 1550$ nm. Listing 1 shows sample Verilog-A model for a waveguide, based on [12]. The

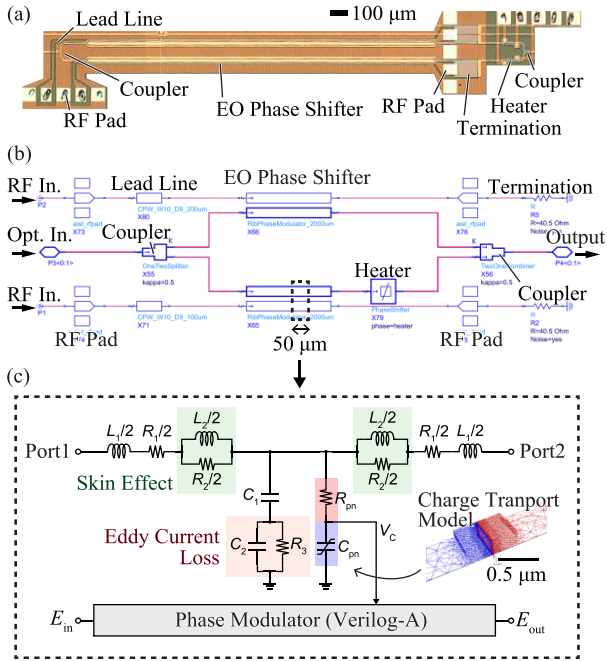
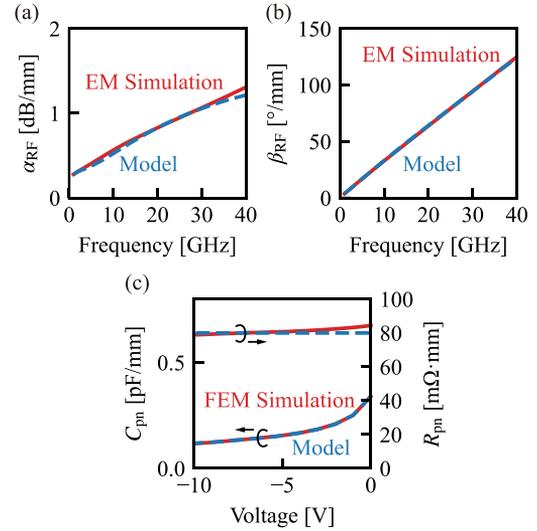

FIGURE 2. Modeling of Si rib-waveguide Mach-Zehnder modulator. (a) Optical micrograph of fabricated device. (b) Model schematic. (c) Subcircuit of the segment of the traveling-wave electrodes.

FIGURE 3. Fitting of the electrical parameters. (a) RF loss constant. (b) RF phase constant. (c) Capacitance and resistance of the p-n junction.

Polar2Cartesian() function converts polar coordinates to Cartesian coordinates, and the CartesianMultiplier() function computes the product of complex numbers, as described in Appendix 1. The group delay expressed in (1) is implemented by the absolute delay operator absdelay() predefined in Verilog-A.

### B. Coupler

The couplers are used to split, combine, and loopback the light. Different models are used to match the forward and backward propagation definitions of the input and output ports for each application. The one-by-two couplers are dominated by

$$\begin{pmatrix}\tilde{E}_{o1}\\ \tilde{E}_{o2}\end{pmatrix} = \begin{pmatrix}\sqrt{\kappa}\\ \sqrt{1-\kappa}\end{pmatrix}\tilde{E}_{in}, \quad (4)$$

$$\tilde{E}_{out} = (\sqrt{\kappa} \quad \sqrt{1-\kappa}) \cdot \begin{pmatrix}\tilde{E}_{i1}\\ \tilde{E}_{i2}\end{pmatrix}, \quad (5)$$

where $\kappa$ is the coupling power ratio. See Listing 2 and 3 for sample codes of these models. For a two-by-two (directional) coupler, which is dominated by [31]

$$\begin{pmatrix}\tilde{E}_{o1}\\ \tilde{E}_{o2}\end{pmatrix} = \begin{pmatrix}\sqrt{1-\kappa} & -j\sqrt{\kappa}\\ -j\sqrt{\kappa} & \sqrt{1-\kappa}\end{pmatrix} \cdot \begin{pmatrix}\tilde{E}_{i1}\\ \tilde{E}_{i2}\end{pmatrix}, \quad (6)$$

the couplers typically have an excess loss of about 0.5 dB. This loss is modeled by connecting the optical attenuator (Listing 4) to the input and output ports of the coupler.

### C. Rib Waveguide Mach-Zehnder Modulator

Mach-Zehnder optical modulators are particularly difficult to model because they require accurate knowledge of RC time constants, RF losses, optical and electrical group delays, and characteristic impedance of the electrode. For this reason, most use cases of the Verilog-A model of Mach-Zehnder modulators are limited up to 25 Gbaud due to low accuracy in fitting RF characteristics [25, 26]. Fig. 2(a) and (b) show the optical micrograph and developed model of the rib-type Mach-Zehnder modulator. The input light is split at the coupler and path through the EO phase shifter. A thermo-optic heater adjusts the bias phase between the arms. The RF signals are input from the RF pads, propagate through the traveling-wave electrodes of the EO phase shifters, and are terminated with the resistors. The equivalent circuit model of the EO phase shifter is shown in Fig. 2(c). In this study, we incorporated the skin effect in the metal wiring and eddy currents in the substrate into the model. $L_1$ and $R_1$ are the wiring inductance and resistance, respectively. $L_2$ and $R_2$ model the increase in parasitic resistance at high frequencies due to the skin effect. $C_1$ represents the substrate capacitance, and $C_2$ and $R_3$ model the reduction in shunt impedance at high frequencies due to eddy currents. These parameters were extracted by electromagnetic analysis using Keysight's Momentum. $R_{pn}$ and $C_{pn}$ are the junction resistor and capacitor at the pn junction, which were extracted by charge transport simulation using Ansys's Lumerical. $C_{pn}$ has a voltage dependency due to the expansion of the depletion layer, which is modeled by a polynomial function for the applied voltage $V_C$. The phase shifter is driven by

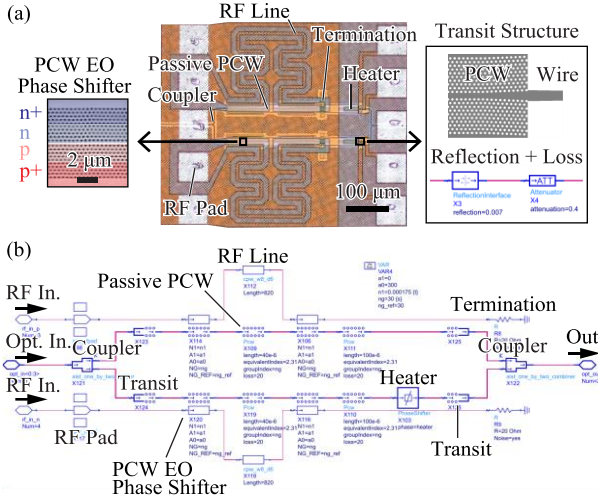

**FIGURE 4.** Modeling of a Si PCW Mach-Zehnder modulator. (a) Micrograph of the modulator and transition structure between the PCW and thin wire. The transition was modeled by optical reflections and attenuators (loss). (b) Model schematic. Similar to the rib type, the EO phase shifter model consists of subcircuits of the segmented traveling-wave electrode.

$V_C$, changing the mode equivalent refractive index $n_{eq}$ and optical loss $\alpha$ as follows:

$$n_{eq} = n_0 + n_1 V_{in} + n_2 V_{in}^2 \cdots \quad (7)$$

$$\alpha = \alpha_0 + \alpha_1 V_{in} + \alpha_2 V_{in}^2 \cdots \quad (8)$$

The traveling-wave electrodes with a length of 2.0 mm were modeled by connecting 40 segments with a length of 50 μm in series. This distributed model accurately simulates the RF propagation losses, impedance mismatch, and phase mismatch between the optical and RF signals. Fig. 3(a) and (b) show the fitting results of the RF propagation constants of the electrodes, $\gamma_{RF} = \alpha_{RF} + j\beta_{RF}$, which can be calculated from the S-parameters as follows:

$$e^{-\gamma_{RF} l} = \left(\frac{1 - S_{11}^2 + S_{21}^2}{2 S_{21}} \pm K\right)^{-1} \quad [32], \quad (9)$$

By properly accounting for skin effects and eddy current losses, good fits were obtained up to 40 GHz for $\gamma_{RF}$. In addition, the voltage dependency of $C_{pn}$ is properly modeled as shown in Fig. 3(c), which enables accurate simulation of the frequency response.

### C. PCW Mach-Zehnder Modulator

Fig. 4(a) and (b) show the optical micrograph and model of the Si PCW Mach-Zehnder modulator. The slow light effect generated in the PCW enhances the light-matter interaction, and the phase shift in the PCW, $\Delta\varphi$, is increased in proportion to $n_g$, as follows:

$$\Delta\phi = k_0 n_g \frac{\Delta n_{eq}}{n_{eq}} \zeta l_{pcw}, \quad [33] \quad (10)$$

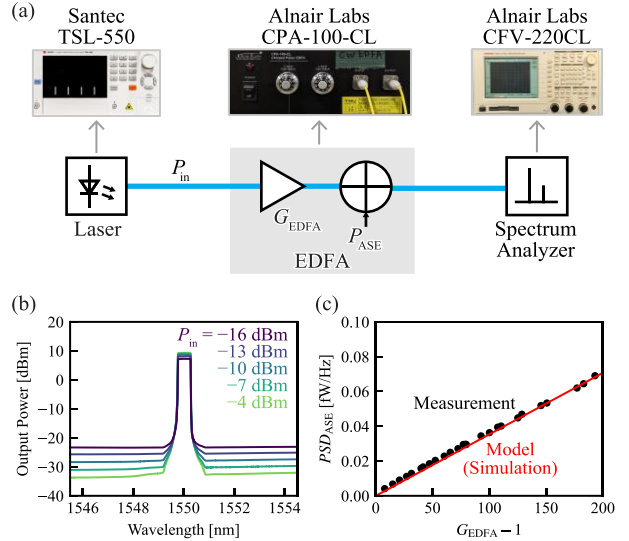

**FIGURE 5.** Modeling of an EDFA. (a) Experimental setup for noise modeling of the EDFA and block diagram of the model. (b) Measured power spectrum output from the EDFA. (c) Measured and simulated power spectral densities of the ASE.

where $\Delta n_{eq}$, $l_{pcw}$, and $\zeta$ represent the change of $n_{eq}$, the phase shifter length, and a constant value, respectively. Listing 5 shows the model of the PCW phase shifter. In this model, the first-order voltage coefficient of the refractive index, $n_1$, was replaced by $(n_g/n_{g\_ref})n_1$ to simulate the slow light enhancement, where $n_{g\_ref}$ is $n_g$ without significant slow light effect. To improve the phase matching between slow light and RF signals, the phase shifter was divided into two segments, and RF delay lines were inserted between them [34]. The RF lines were modeled by using the transmission line predefined in ADS. Undoped passive PCWs are placed between the phase shifter segments to electrically separate each segment. To properly account for the optical delay due to the passive PCWs, we inserted a waveguide into the model. The model parameters were extracted based on device simulations in the same manner as those for the rib-waveguide modulator. Optical reflectors (Listing 6) with a reflectivity of 7.5% were inserted at the input and output of the PCW, and the connection structure was modeled.

### III. TESTING EQUIPMENT

In this section, we develop models for testing equipment to construct a receiver composed of an EDFA, tunable filter, and photodetector module. Our model introduces noise, which has a significant impact on the bit error rate and has been overlooked in traditional EO co-simulation. Here we measure and model the noise of each device used in our lab to improve the accuracy of our simulations.

### A. EDFA

EDFAs compensate for transmission losses and increase receiver sensitivity, but amplified spontaneous emission

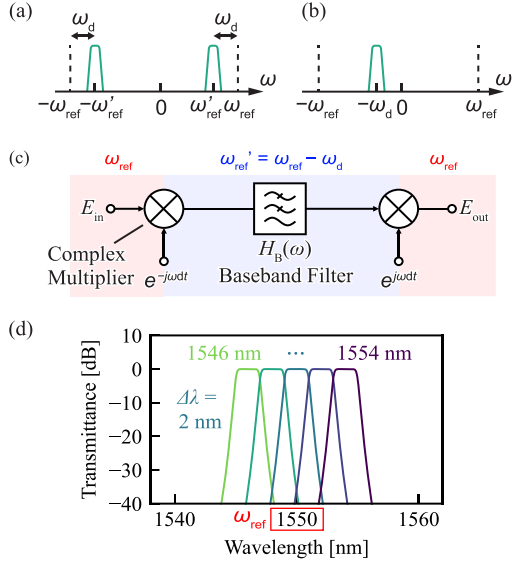

**FIGURE 6.** Modeling of a tunable filter. (a) Frequency response of optical bandpass filter. (b) Frequency response of frequency-shifted equivalent filter. (c) Block diagram of the model. (d) Simulated transmission spectrum.

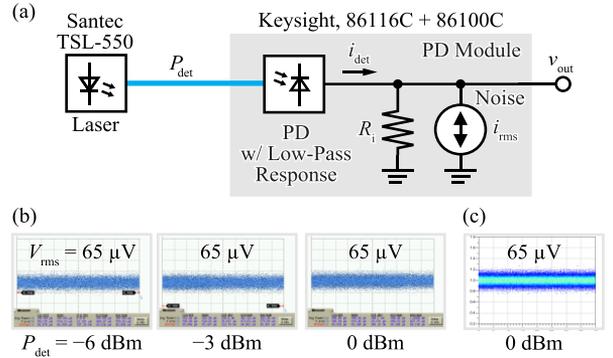

**FIGURE 7.** Modeling of a photodetector module. (a) Experimental setup for noise modeling of the module and block diagram of the model. (b) Measured noise waveform of Keysight's 86116C photodetector module. (c) Simulated noise waveform.

(ASE) introduces noise that must be modeled. The input power is amplified by $G_{EDFA}$ and then ASE approximated by Gaussian noise is added. The noise power level, $P_{ASE}$, is given by

$$P_{ASE} = \mu h\nu \Delta\nu (G_{EDFA} - 1) \quad (11)$$

where $\mu$ is the inversion factor, $h\nu$ is the photon energy, and $\Delta\nu$ is the frequency bandwidth [35]. Fig. 5(a) and (b) show the measurement setup and measured output spectrum of an EDFA (Alnair Labs' CPA-100-CL). When the output power is controlled at a constant value, a smaller $P_{in}$ results in a larger $G_{EDFA}$ and a larger $P_{ASE}$. The power spectral density, $PSD_{ASE}$, can be calculated from the wavelength resolution of the spectrum analyzer, $\lambda_{res} = 0.5$ nm. Fig. 5(c) compares the measurement and model of $PSD_{ASE}$ for $(G_{EDFA} - 1)$. The coefficient $\mu h\nu \Delta\nu$ was determined to be $3.44 \times 10^{-19}$ W/Hz by the least-squares method. The noise bandwidth in the simulation is equal to the sampling frequency of the simulation, i.e. $\Delta\nu = f_s$. Listing 7 shows the Verilog-A code of the EDFA. A Gaussian random $rdist\_normal()$ function was used for noise generation, with the standard deviation set to be the square root of the noise power.

### B. Tunable Filter

Optical filters are used to remove ASE generated by the EDFA. As mentioned earlier, the optical frequency is shifted to the baseband by $\omega_{ref}$ to reduce the number of time steps in the transient analysis. Consequently, the frequency response of the filter must also be shifted accordingly. Fig. 6(a) and (b) show the responses of the bandpass filter before and after the frequency shift, respectively. The center frequency of the filter must be flexibly set to any value other than $\omega_{ref}$. We express the difference between $\omega_{ref}$ and the center frequency of the filter as $\omega_d$. This causes the frequency response of the baseband-equivalent filter to become asymmetric around 0 Hz, resulting in complex coefficients in the Laplace transfer function. Since Verilog-A does not natively support complex numbers, our model locally adjusts the reference frequency. Fig. 6(c) shows a block diagram of the tunable filter. In the time domain, the frequency shift corresponds to a multiplication by $\exp(-j\omega_d t)$, so the tunable filter is implemented by inserting a complex multiplier before and after the baseband low-pass filter. Listing 8 shows the Verilog-A code for the tunable filter. The center wavelength is specified by the parameter 'wavelength'. 'osc1' and 'osc2' serve as local oscillators for converting the reference frequency at the input and output. The baseband low-pass filter was implemented using the laplace_nd() function predefined in Verilog-A. The coefficients of the Laplace transfer function were determined by a MATLAB script and are provided as an array. The filter response was approximated using a sixth-order Butterworth filter, which offers a flat passband with minimal impact on the signal and sufficient roll-off to eliminate ASE. The wavelength bandwidth, $\Delta\lambda$, was set to 2 nm. Fig. 6(d) shows the simulated transmission spectrum. The center wavelength was varied from 1546 nm to 1554 nm in 2 nm increments, with $\lambda_{ref}$ fixed at 1550 nm, demonstrating the wavelength tuning operation.

### C. Photo Detector Module

The photodetector module consists of a photodiode (PD), a trans-impedance amplifier (TIA), and an analog-to-digital converter (ADC). The PD outputs a current $i_{det} = RP_{det}$, with the received power $P_{det}$ and the responsivity $R$. The RC time constant at the PD was introduced into the model by a first-order low-pass transfer function using the laplace_nd() function. In addition, we introduced the thermal noise from the TIA and ADC. Fig. 7(a) shows the measurement setup and corresponding model of the PD module (Keysight's 86116C). The TIA was replaced by a load resistor $R_L$. The TIA noise and ADC noise are modeled by the thermal noise generated from $R_L$. The root-mean-squared (RMS) voltage of the thermal noise, $V_{rms}$, is given by

$$V_{rms} = \sqrt{4kTR_L B} \qquad (12)$$

where $k$, $T$, and $B$ represent the Boltzmann constant, absolute temperature, and bandwidth, respectively. Fig. 7(b) shows the measured noise waveforms when $P_{det}$ was varied from $-\infty$ to 0 dBm. The measured RMS voltage was $V_{rms} = 65\ \mu V$. The fact that $V_{rms}$ is independent of $P_{det}$ suggests that the primary sources of noise are the TIA and ADC, rather than the shot noise of the photodiode. From Eq. (12), $R_L$ was determined to be 0.1 Ω to produce thermal noise with $V_{rms} = 65\ \mu V$. Here, $T = 27°C$ and $B = 2500$ GHz were used by setting parameters "Temp" and "NoiseBandwidth" in the resistor model of ADS. $R$ was set to 10 so that the output power is directly displayed as a voltage. As shown in Fig. 7(c), the simulated noise amplitude was $V_{rms} = 65\ \mu V$, which was consistent with the measurement.

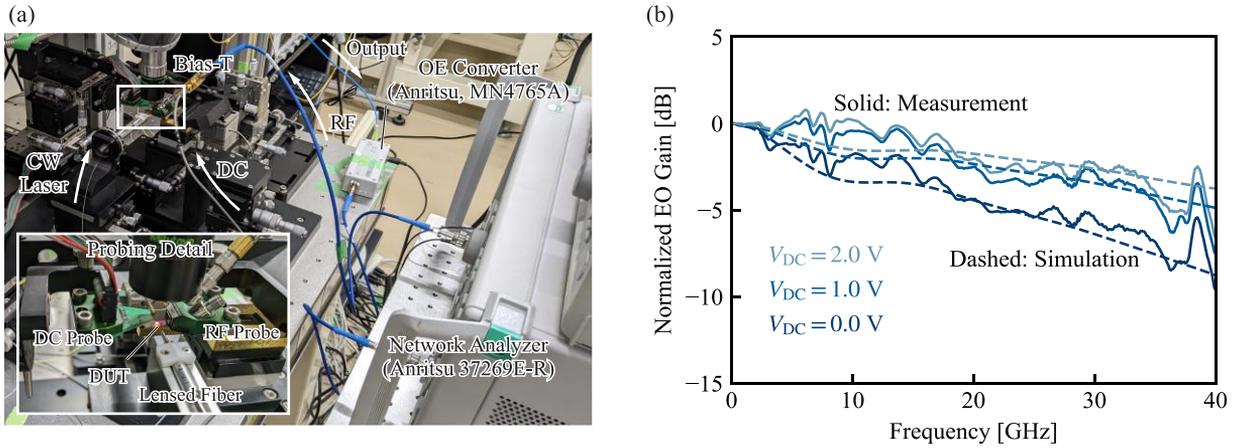

**FIGURE 8.** Experimental validation of the Si optical modulator chip. (a) Experimental setup for S-parameter measurement. (b) Simulated EO gain of rib waveguide modulator with different bias voltages.

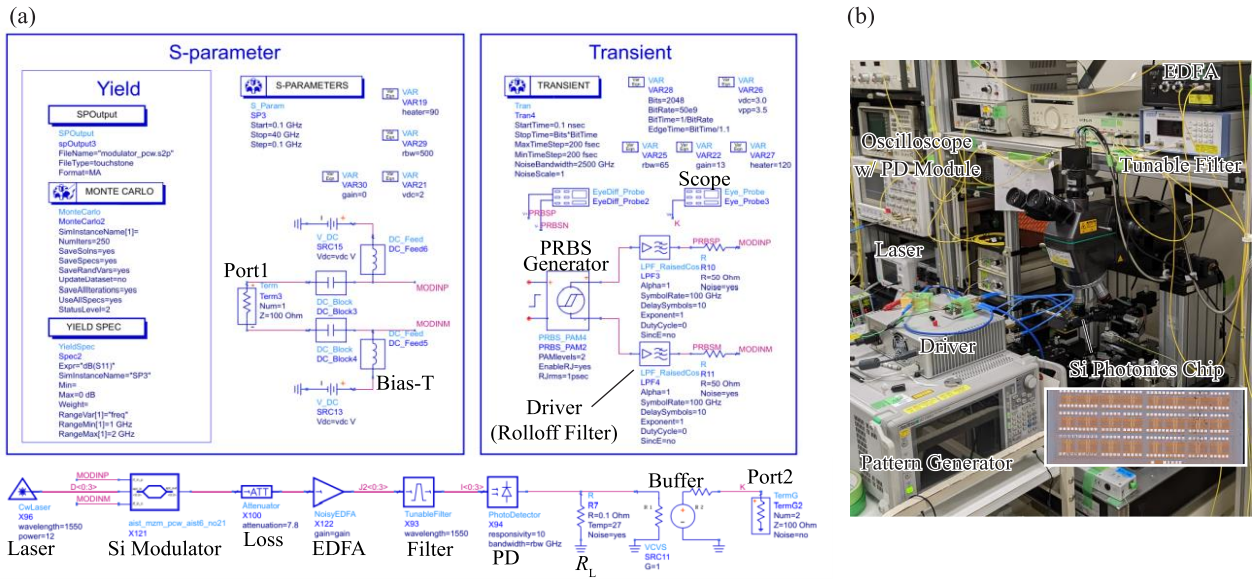

**FIGURE 9.** Experimental validation of the full optical link. (a) Test bench for signal transmission constructed in ADS. (b) Experimental setup for >50-Gbaud signal transmission.

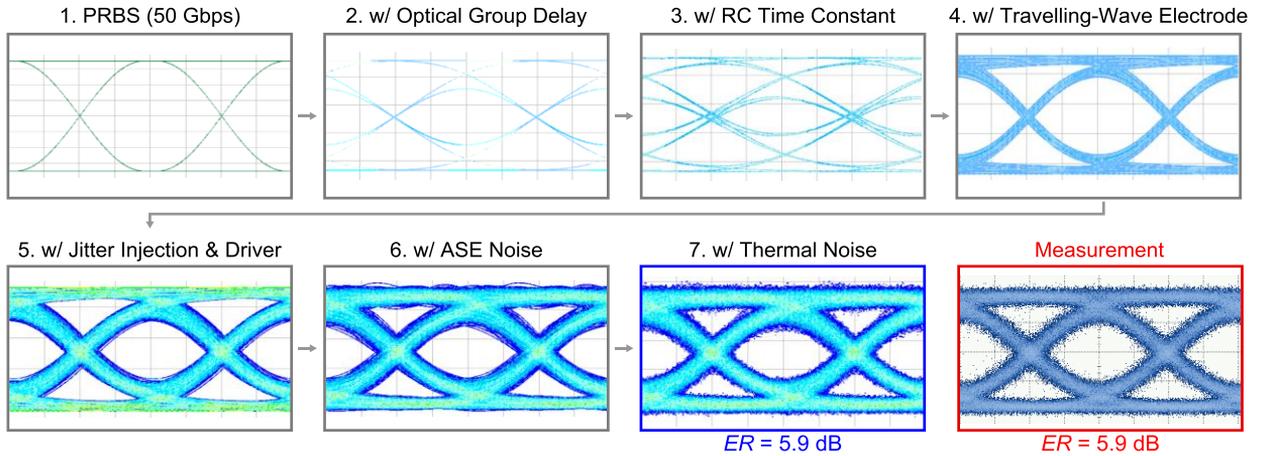

**FIGURE 10.** Simulated and measured eye diagrams for the rib-type Mach-Zehnder modulator at 50 Gbaud. The simulated eye diagrams show the sequential changes due to the addition of physical properties.

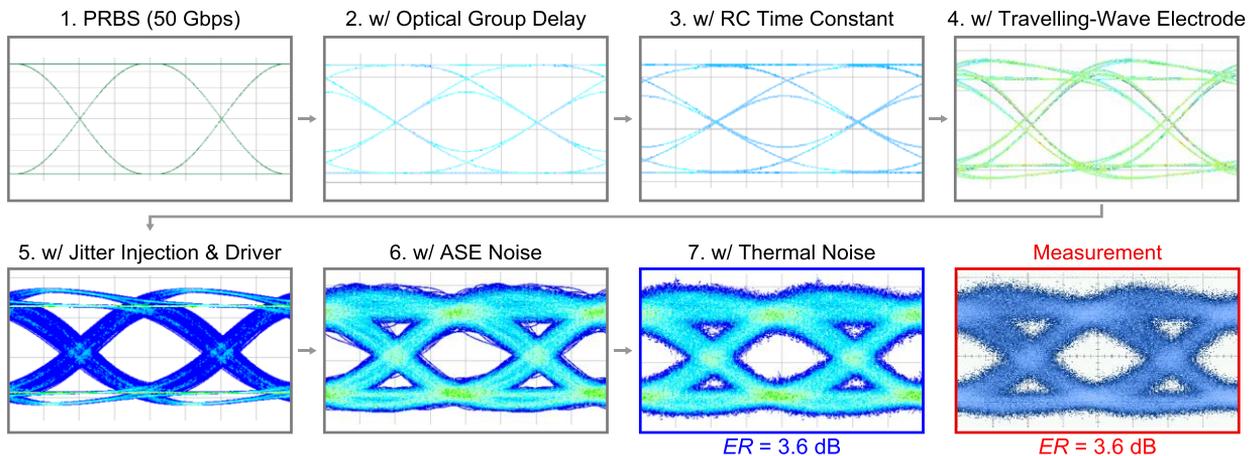

**FIGURE 11.** Simulated and measured eye diagrams for the PCW-type Mach-Zehnder modulator at 50 Gbaud. The simulated eye diagrams show the sequential changes due to the addition of physical properties.

## IV. EXPERIMENTAL VALIDATION

Firstly, we evaluated the EO S parameters of the modulator using a network analyzer (Anritsu, 37269E-R) and a photodetector module (Anritsu, MN4765A) as shown in Fig. 8(a). Fig. 8(b) shows the measured and simulated frequency responses of the rib-waveguide modulator for various bias voltages $V_{DC}$. The simulated responses (dashed lines) agreed roughly with measurements (solid lines) from DC to 40 GHz thanks to the inclusion of RF loss, EO phase mismatch, impedance mismatch, and RC time constant. The voltage dependence was also well simulated by considering the nonlinearity of $C_{pn}$.

Then, we built a simulation test bench corresponding to the experimental setup using the developed model library and verified the accuracy of the models as shown in Fig. 9. CW light output from a tunable laser source (Santec's TSL-550) was input to a Si optical modulator. The modulated light passed through an EDFA (Alnair Labs' CPA-100-CL) and bandpass filter (Alnair Labs' CVF-220CL). The eye diagrams were observed by a sampling oscilloscope equipped with a photodetector module (Keysight's 86116C, 86100C) while driving the modulator using a broadband amplifier (SHF's S804B). The pseudo-random bit sequence (PRBS) was generated by a pulse pattern generator (Anritsu's MP1800A) and a multiplexer (SHF's 601A). The RF loss in the cable was compensated by an equalizer (SHF's EQ25A). The modulators were push-pull driven at $V_{pp}$ = 3.5 V. $P_{LD}$ and $P_{det}$ were set to 12 dBm and 5 dBm, respectively. In the simulation, all the components were replaced with our Verilog-A models or built-in electrical models in the simulator. The driver was replaced by a roll-off filter with a bandwidth of 50 GHz, and an RMS jitter of 1 ps was injected into the PRBS signal source, corresponding to the measurement. The modulation signals were 50- and 64-Gbps non-return-to-zero PRBS signals, and 2048 bits were transmitted at a fixed time step of 200 fs. The first 0.1 ns of the waveform was discarded to eliminate transients at the beginning of the simulation.

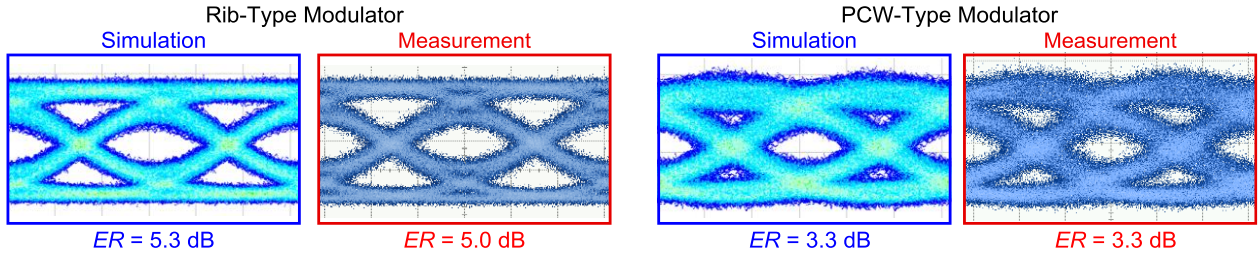

FIGURE 12. Simulated and measured eye diagrams for the rib- and PCW-type Mach-Zehnder modulator at 64 Gbaud.

Fig. 10 and 11 show the results of 50-Gbaud signal transmission using ribbed- and PCW-type modulators, respectively. In contrast to previously reported Verilog-A-based simulations, our simulations introduced physical properties that have a noticeable impact on signal quality in modulators and test equipment. These eye diagrams show the transitions as the various physical properties described in the modeling section were added. The optical group delay and the RC time constant of the modulator degrade the eye-opening. The traveling wave electrode improves eye-opening by matching the optical signal with the RF signal. Furthermore, the proper modeling of jitter and noise accurately simulated the measured waveform. Fig. 12 shows the measured and simulated eye diagrams at 64 Gbaud. The simulated results agreed well with the measurements under all conditions, with an error in extinction ratio (ER) of less than 1 dB. Therefore, the developed models enable precise prediction of transceiver link performance at high data rates exceeding 50 Gbaud, indicating its potential for designing 800G, 1.6T, and beyond.

## V. CONCLUSION

In this study, we developed accurate and interoperable photonic device models that reflect the physical properties of basic Si photonics devices and equipment. We incorporated RF losses, optical losses, EO phase mismatch, impedance mismatch, and RC time constant into the models of Si rib waveguide and PCW Mach-Zehnder modulators. The S-parameter simulations showed good agreement with measurements up to 40 GHz, accurately predicting voltage dependence and process variations. In the signal transmission simulation, we incorporated all signal degradation factors present in the testing equipment, including jitter, ASE, and thermal noise. Consequently, the simulations accurately matched measurements at 50 Gbps and higher. These results facilitate precise verification and performance prediction of EO integrated circuits, paving the way for emerging applications.

## APPENDIX

The key models that construct the testbench are shown in Listing 1 to 8. See the GitHub repository for the complete model library [29].

```
1: module Waveguide(node1, node2);
2:     inout [0:3] node1;
3:     inout [0:3] node2;
4:     optical [0:3] node1, node2;
5:     optical [0:1] fwd, bwd, transfer_pol, transfer;
6:
7:     parameter real length = 100e-6; // [m]
8:     parameter real equivalentIndex = 2.31;
9:     parameter real groupIndex = 4.34;
10:    parameter real loss = 2.0; // [dB/cm]
11:    real alpha = 23.0258509299404568 * loss;
12:
13:    Polar2Cartesian Polar2Cartesian1(transfer_pol, transfer);
14:
15:    // Forward and backward propagation
16:    CartesianMultiplier CartesianMultiplier1(transfer, node1[0:1], fwd[0:1]);
17:    CartesianMultiplier CartesianMultiplier2(transfer, node2[2:3], bwd[0:1]);
18:
19:    analog begin
20:        // Propagation loss and phase rotation
21:        OptE(transfer_pol[0]) <+ exp(-alpha / 2 * length);
22:        OptE(transfer_pol[1]) <+ (-length * equivalentIndex * 2 * `M_PI * `F_REF / `P_C) % (2 * `M_PI);
23:
24:        // Output
25:        OptE(node2[0]) <+ absdelay(OptE(fwd[0]), length * groupIndex / `P_C);
26:        OptE(node2[1]) <+ absdelay(OptE(fwd[1]), length * groupIndex / `P_C);
27:        OptE(node1[2]) <+ absdelay(OptE(bwd[0]), length * groupIndex / `P_C);
28:        OptE(node1[3]) <+ absdelay(OptE(bwd[1]), length * groupIndex / `P_C);
29:    end
30: endmodule
```

LISTING 1. Verilog-A model of the waveguide (Waveguide.va).

```
1: module OneTwoSplitter(one, two1, two2); // one[0:1]:input, one[2:3]:output, two1[0:1]:output, two1[2:3]:input, two2[0:1]:output, two2[2:3]:input
```

```
2:      inout [0:3] one, two1, two2;
3:      optical [0:3] one, two1, two2;
4:
5:      parameter real kappa = 0.5; // Power ratio
6:
7:      analog begin
8:          OptE(two1[0]) <+ sqrt(kappa) * OptE(one[0]);
9:          OptE(two1[1]) <+ sqrt(kappa) * OptE(one[1]);
10:         OptE(two2[0]) <+ sqrt(1 - kappa) * OptE(one[0]);
11:         OptE(two2[1]) <+ sqrt(1 - kappa) * OptE(one[1]);
12:         OptE(one[2]) <+ OptE(two1[2]) * sqrt(kappa) + OptE(two2[2]) * sqrt(1 - kappa);
13:         OptE(one[3]) <+ OptE(two1[3]) * sqrt(kappa) + OptE(two2[3]) * sqrt(1 - kappa);
14:     end
15: endmodule
```

**LISTING 2.** Verilog-A model of the one-by-two splitter (OneTwoSplitter.va).

```
1: module TwoOneCombiner(two1, two2, one);
2:      inout [0:3] two1, two2, one;
3:      optical [0:3] two1, two2, one;
4:
5:      parameter real kappa = 0.5; // Power ratio
6:
7:      analog begin
8:          OptE(one[0]) <+ OptE(two1[0]) * sqrt(kappa) + OptE(two2[0]) * sqrt(1 - kappa);
9:          OptE(one[1]) <+ OptE(two1[1]) * sqrt(kappa) + OptE(two2[1]) * sqrt(1 - kappa);
10:         OptE(two1[2]) <+ sqrt(kappa) * OptE(one[2]);
11:         OptE(two1[3]) <+ sqrt(kappa) * OptE(one[3]);
12:         OptE(two2[2]) <+ sqrt(1 - kappa) * OptE(one[2]);
13:         OptE(two2[3]) <+ sqrt(1 - kappa) * OptE(one[3]);
14:     end
15: endmodule
```

**LISTING 3.** Verilog-A model of the one-by-two combiner (TwoOneCombiner.va).

```
1: module Attenuator(in, out);
2:      inout [0:3] in, out;
3:      optical [0:3] in, out;
4:      optical [0:1] transfer_pol, transfer;
5:
6:      parameter real attenuation = 3.0; // [dB]
7:
8:      Polar2Cartesian Polar2Cartesian1(transfer_pol, transfer);
9:      CartesianMultiplier CartesianMultiplier1(transfer, in[0:1], out[0:1]);
10:     CartesianMultiplier CartesianMultiplier2(transfer, out[2:3], in[2:3]);
11:
12:     analog begin
13:         OptE(transfer_pol[0]) <+ pow(10, - attenuation / 20);
14:         OptE(transfer_pol[1]) <+ 0;
15:     end
16: endmodule
```

**LISTING 4.** Verilog-A model of the optical attenuator (Attenuator.va).

```
1: module PcwPhaseModulator(opt_in, opt_out, ele_in);
2:      inout [0:3] opt_in, opt_out; // Optical input and output
3:      input ele_in; // Electrical input
4:      optical [0:3] opt_in, opt_out, transfer_pol, transfer, out_tmp;
5:      optical [0:1] fwd, bwd, transfer_pol, transfer;
6:      electrical ele_in;
7:
8:      parameter real length = 100e-6; // [m]
9:      parameter real equivalentIndex = 2.31; // at 0-V bias
10:     parameter real groupIndex = 20;
11:     parameter real refGroupIndex = 20;
12:     parameter real loss = 2.0; // [dB/cm] at 0-V bias
13:
14:     real n0 = equivalentIndex;
15:     parameter real n1 = 0.0002;
16:     real a0 = 23.0258509299404568 * loss;
17:     parameter real a1 = 10;
18:
19:     Polar2Cartesian Polar2Cartesian1(transfer_pol, transfer);
20:     CartesianMultiplier CartesianMultiplier1(transfer, opt_in[0:1], fwd);
21:     CartesianMultiplier CartesianMultiplier2(transfer, opt_out[2:3], bwd);
22:
23:     real polynomialEquivalentIndex;
24:     real polynomialAlpha;
25:
26:     analog begin
27:         // Voltage dependency
28:         polynomialEquivalentIndex = n0 + n1 * groupIndex / refGroupIndex * V(ele_in); // Slow-light enhancement
29:         polynomialAlpha = a0 + a1 * V(ele_in);
30:
31:         // Transfer function
32:         OptE(transfer_pol[0]) <+ exp(- polynomialAlpha / 2 * length); // Propagation loss
33:         OptE(transfer_pol[1]) <+ (- length * polynomialEquivalentIndex * 2 * `M_PI * `F_REF / `P_C) % (2 * `M_PI); // Phase rotation
34:
35:         // Output
36:         OptE(opt_out[0]) <+ absdelay(OptE(fwd[0]), length*groupIndex / `P_C);
37:         OptE(opt_out[1]) <+ absdelay(OptE(fwd[1]), length*groupIndex / `P_C);
38:         OptE(opt_in[2]) <+ absdelay(OptE(bwd[0]), length*groupIndex / `P_C);
39:         OptE(opt_in[3]) <+ absdelay(OptE(bwd[1]), length*groupIndex / `P_C);
40:     end
41: endmodule
```

**LISTING 5.** Verilog-A model of the PCW phase shifter (PcwPhaseModulator.va).

```
1: module ReflectionInterface(node1, node2);
2:      inout [0:3] node1;
3:      inout [0:3] node2;
4:      optical [0:3] node1, node2;
5:
6:      parameter real reflection = 0;
7:
8:      analog begin
9:          OptE(node1[2]) <+ sqrt(1 - reflection) * OptE(node2[2]) + sqrt(reflection) * OptE(node1[0]);
10:         OptE(node1[3]) <+ sqrt(1 - reflection) * OptE(node2[3]) + sqrt(reflection) * OptE(node1[1]);
11:         OptE(node2[0]) <+ sqrt(1 - reflection) * OptE(node1[0]) + sqrt(reflection) * OptE(node2[2]);
12:         OptE(node2[1]) <+ sqrt(1 - reflection) * OptE(node1[1]) + sqrt(reflection) * OptE(node2[3]);
13:     end
14: endmodule
```

**LISTING 6.** Verilog-A model of optical reflection (ReflectinInterface.va).

```
1: module NoisyEDFA(in, out);
2:      inout [0:3] in, out;
3:      optical [0:3] in, out;
4:      optical [0:1] transfer, transfer_pol, noise_pol, out_tmp, out_tmp_pol, out_pol;
5:
6:      parameter real gain = 3.0; // [dB]
7:      integer seed = 52924;
8:      integer mean = 0;
9:      real noise_density = 0; // [W/Hz]
10:     real noise_power = 0; // [W]
11:     real fs = 5e12;     // sample rate [Hz]
12:
13:     // EDFA Gain
14:     Polar2Cartesian Polar2Cartesian1(transfer_pol, transfer);
15:     CartesianMultiplier CartesianMultiplier1(transfer, in[0:1], out_tmp);
16:
17:     // ASE Noise
18:     Cartesian2Polar Cartesian2Polar1(out_tmp, out_tmp_pol);
19:     CartesianAdder CartesianAdder1(out_tmp_pol, noise_pol, out_pol);
20:     Polar2Cartesian Polar2Cartesian2(out_pol, out[0:1]);
21:
22:     analog begin
23:         @(initial_step) begin
24:             noise_density = 3.44e-19 * (pow(10, gain / 10) - 1);
25:             noise_power = noise_density * fs;
26:         end
27:
28:         OptE(noise_pol[0]) <+ $rdist_normal(seed, mean, sqrt(noise_power)); // ASE Amplitude
29:         OptE(noise_pol[1]) <+ 0; // No phase noise
30:         OptE(transfer_pol[0]) <+ pow(10, gain / 20); // Gain
31:         OptE(transfer_pol[1]) <+ 0; // No phase rotation
32:
33:         // Ideal isolation
34:         OptE(in[2]) <+ 0;
35:         OptE(in[3]) <+ 0;
36:     end
37: endmodule
```

**LISTING 7.** Verilog-A model of an EDFA (NoisyEDFA.va).

```
1: module TunableFilter(in, out);
2:      inout [0:3] in, out;
3:      optical [0:3] in, out, in_tmp, out_tmp;
4:      optical [0:1] osc1, osc2;
5:
6:      parameter real wavelength = 1551; // [nm]
7:      real center_freq = `P_C / wavelength / 1e-9; // [Hz]
8:
9:      // Reference frequency conversion
10:     CartesianMultiplier CartesianMultiplier1(in[0:1], osc1, in_tmp[0:1]);
11:     CartesianMultiplier CartesianMultiplier2(out_tmp[0:1], osc2, out[0:1]);
12:
13:     CartesianMultiplier CartesianMultiplier3(out[2:3], osc1, out_tmp[2:3]);
14:     CartesianMultiplier CartesianMultiplier4(in_tmp[2:3], osc2, in[2:3]);
15:
16:     analog begin
17:         // delta=2nm
18:         OptE(out_tmp[0]) <+ laplace_nd(OptE(in_tmp[0]), {2.347141585877207e+71,0,0,0,0,0,0}, {2.347141585877208e+71,1.154657487839621e+60,2.840122475453116e+48,4.428868818445329e+36,4.604233134433859e+24,3.034545479782387e+12,1});
19:         OptE(out_tmp[1]) <+ laplace_nd(OptE(in_tmp[1]), {2.347141585877207e+71,0,0,0,0,0,0}, {2.347141585877208e+71,1.154657487839621e+60,2.840122475453116e+48,4.428868818445329e+36,4.604233134433859e+24,3.034545479782387e+12,1});
20:         OptE(in_tmp[2]) <+ laplace_nd(OptE(out_tmp[2]), {2.347141585877207e+71,0,0,0,0,0,0}, {2.347141585877208e+71,1.154657487839621e+60,2.840122475453116e+48,4.428868818445329e+36,4.604233134433859e+24,3.034545479782387e+12,1});
21:         OptE(in_tmp[3]) <+ laplace_nd(OptE(out_tmp[3]), {2.347141585877207e+71,0,0,0,0,0,0}, {2.347141585877208e+71,1.154657487839621e+60,2.840122475453116e+48,4.428868818445329e+36,4.604233134433859e+24,3.034545479782387e+12,1});
22:
23:         // Local oscillation for the reference frequency converison
24:         OptE(osc1[0]) <+ cos(-2 * `M_PI * (center_freq - `F_REF) * $abstime);
25:         OptE(osc1[1]) <+ sin(-2 * `M_PI * (center_freq - `F_REF) * $abstime);
26:         OptE(osc2[0]) <+ cos(2 * `M_PI * (center_freq - `F_REF) * $abstime);
27:         OptE(osc2[1]) <+ sin(2 * `M_PI * (center_freq - `F_REF) * $abstime);
28:     end
29: endmodule
```

**LISTING 8.** Verilog-A model of the tunable filter (TunableFilter.va).

## ACKNOWLEDGMENT

This work was supported through the activities of VDEC, The University of Tokyo, in collaboration with Keysight Technologies, Inc. and Cadence Design Systems, Inc. We thank Dr. Mikiya Kamata for the modulator design and Dr. Takemasa Tamanuki for technical advice on the EDFA evaluation.

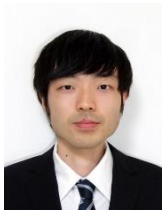

**KEISUKE KAWAHARA** (Graduate Student Member, IEEE) received the B.E. and M.E. degrees in engineering from Tokyo University of Science, Chiba, Japan, in 2020 and 2022, respectively. He is currently pursuing a Ph.D. degree with the Graduate School of Engineering Science, Yokohama National University, Kanagawa, Japan. His research interests include electronic and photonic integrated circuits.

Mr. Kawahara is also a student member of Optica, the Institute of Electronics, Information and Communication Engineers (IEICE), and the Japan Society of Applied Physics (JSAP).

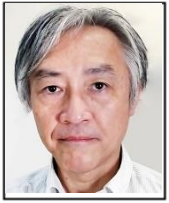

**TOSHIHIKO BABA** (Fellow, IEEE) received the B.E., M.E., and Ph.D. degrees all from the Division of Electrical and Computer Engineering, Yokohama National University in 1985, 1987, and 1990, respectively. He became an associate professor and full professor at this university in 1994 and 2005, respectively. He has studied ARROW waveguides, VCSELs, micro/nano lasers and spontaneous emission control, photonic crystals, Si photonics, slow light, biosensors, optical modulators and LiDAR. He is a Fellow Member of JSAP and Optica, as well as an Associate Member of the Science Council of Japan. He served as a vice president of JSAP from 2018–2020. He was the recipient of the JSPS Prize in 2006, the IEEE/LEOS Distinguished Lecturer in 2006–2007, Commendation for Science and Technology by the MEXT in 2016, and Medal of Purple Ribbon in 2024.